\documentclass[onecolumn,twoside]{IEEEtran}
\usepackage{mathpazo}
\usepackage{times}
\usepackage{color}
\usepackage{amsmath}
\usepackage{amsfonts}
\usepackage{latexsym}
\usepackage{amssymb}
\usepackage{cite}

\usepackage{upref}
\usepackage{theorem}
\usepackage{graphicx}
\usepackage{psfrag}

\usepackage[normalem]{ulem}

\theoremstyle{plain}
\theorembodyfont{\normalfont\slshape}

\newtheorem{thm}{Theorem$\!$}
\newenvironment{theorem}
{\begin{thm}\hspace*{-1ex}{\bf.}}{\end{thm}}

\newtheorem{lem}{Lemma$\!$}
\newenvironment{lemma}{\begin{lem}\hspace*{-1ex}{\bf.}}{\end{lem}}

\newtheorem{prop}{Proposition$\!$}

\newtheorem{cor}{Corollary$\!$}
\newenvironment{corollary}{\begin{cor}\hspace*{-1ex}{\bf.}}{\end{cor}}

\newtheorem{defn}{Definition$\!$}
\newenvironment{definition}{\begin{defn}\hspace*{-1ex}{\bf.}}{\end{defn}}

\newtheorem{xmpl}{Example$\!$}
\newenvironment{example}{\begin{xmpl}\hspace*{-1ex}{\bf.}}{\hfill$\Box$\end{xmpl}}

\newtheorem{cnstr}{Construction$\!$}

\newenvironment{construction}{\begin{cnstr}\hspace*{-1ex}{\bf.}}{\end{cnstr}}

\newcounter{enumrom}
\renewcommand{\theenumrom}{(\roman{enumrom})}

\makeatletter
\renewcommand{\@endtheorem}{\endtrivlist}
\makeatother

\makeatletter
\renewcommand{\thefigure}{{\@arabic\c@figure}}
\renewcommand{\fnum@figure}{{\bf Figure\,\thefigure}}
\makeatother

\newcommand{\cB}{\mathcal{B}}

\newcommand{\cG}{\mathcal{G}}

\newcommand{\cS}{\mathcal{S}}

\newcommand{\cV}{\mathcal{V}}

\newcommand{\mathset}[1]{\left\{#1\right\}}
\newcommand{\abs}[1]{\left|#1\right|}

\newcommand{\parenv}[1]{\left( #1 \right)}

\newcommand{\be}[1]{\begin{equation}\label{#1}}
\newcommand{\ee}{\end{equation}}

\renewcommand{\leq}{\leqslant}

\renewcommand{\geq}{\geqslant}

\renewcommand{\Bbb}{\mathbb}

\newcommand{\Cref}[1]{Co\-ro\-lla\-ry\,\ref{#1}}

\renewcommand{\Bbb}{\mathbb}

\newcommand{\F}{\mathbb{F}}
\newcommand{\Fq}{{{\Bbb F}}_{\!q}}

\newcommand{\N}{{\Bbb N}}

\DeclareMathOperator{\wt}{wt}
\DeclareMathOperator{\rank}{rank}
\DeclareMathOperator{\std}{STD}
\DeclareMathOperator{\rowsp}{rowsp}
\newcommand{\spn}[1]{\left\langle {#1} \right\rangle}
\newcommand{\oc}{\overline{c}}
\newcommand{\og}{\overline{g}}
\newcommand{\ox}{\overline{x}}
\newcommand{\oy}{\overline{y}}
\newcommand{\ou}{\overline{u}}
\newcommand{\ov}{\overline{v}}
\newcommand{\ow}{\overline{w}}
\newcommand{\bc}{\mathbf{c}}
\newcommand{\ozer}{\overline{0}}
\newcommand{\bzer}{\mathbf{0}}
\newcommand{\locn}{r_{\mathrm{n}}}
\newcommand{\locs}{r_{\mathrm{s}}}
\newcommand{\avn}{t_{\mathrm{n}}}
\newcommand{\avs}{t_{\mathrm{s}}}
\newcommand{\sbinom}[2]{\genfrac{[}{]}{0pt}{}{#1}{#2}}
\newcommand{\eqdef}{\triangleq}
\newcommand{\ext}{{\mathrm{ext}}}
\newcommand{\Cpar}{C_{\mathrm{par}}}

\outer\def\proclaim #1. #2\par{\medbreak
 \noindent{\bf#1.\enspace}{\sl#2\par}%
 \ifdim\lastskip<\medskipamount \removelastskip\penalty55\medskip\fi}

\begin{document}

%----------------- The Title Declarations ------------------------------

\title{\textbf{Locality and Availability of Array Codes Constructed from Subspaces}}

\author{
Natalia Silberstein,
Tuvi Etzion,~\IEEEmembership{Fellow,~IEEE},
Moshe Schwartz,~\IEEEmembership{Senior Member,~IEEE}
\thanks{The material in this paper was presented in part at the
  IEEE International Symposium on Information
  Theory (ISIT 2017), Aachen, Germany, June 2017.}%
\thanks{Natalia Silberstein
  was with the Department of Electrical and Computer Engineering, Ben-Gurion University of the Negev, Beer Sheva 8410501, Israel, and is now with Yahoo! Research, Haifa 31905, Israel, (e-mail: natalys@cs.technion.ac.il).}%
\thanks{Tuvi Etzion is with the Department of Computer Science, Technion -- Israel Institute of Technology, Haifa 3200003, Israel, (e-mail: etzion@cs.technion.ac.il).}%
\thanks{Moshe Schwartz is with the Department
   of Electrical and Computer Engineering, Ben-Gurion University of the Negev,
   Beer Sheva 8410501, Israel
   (e-mail: schwartz@ee.bgu.ac.il).}%
\thanks{This work was supported in part by the NSF-BSF grant no.~2016692, and by the Israeli Science Foundation (ISF), Jerusalem, Israel, under Grant no.~130/14.}
}

\maketitle

\begin{abstract}
  We study array codes which are based on subspaces of a linear space
  over a finite field, using spreads, $q$-Steiner systems, and
  subspace transversal designs. We present several constructions of
  such codes which are $q$-analogs of some known block codes such as
  the Hamming and simplex codes. We examine the locality and
  availability of the constructed codes. In particular we distinguish
  between two types of locality and availability -- node vs.~symbol,
  locality and availability. The resulting codes have distinct
  symbol/node locality/availability, allowing a more efficient repair
  process for a single symbol, compared with the repair process for
  the whole node.
\end{abstract}


%%%%%%%%%%%%%%%%%%%%%%%%%%%%%%%%%%%%%%%%%%%%%%%%%%%%%%%%%%%%%%%%%%%%%%%%
%%%%%%%%%%%%%%%%%%%%%%%%%%%%%%%%%%%%%%%%%%%%%%%%%%%%%%%%%%%%%%%%%%%%%%%%
%%%%%%%%%%%%%%%%%%%%%%%%%%%%%%%%%%%%%%%%%%%%%%%%%%%%%%%%%%%%%%%%%%%%%%%%

\section{Introduction}
\IEEEPARstart{D}{esigning} efficient mechanisms to store, maintain,
and efficiently access large volumes of data is a highly relevant
problem.  Indeed, ever-increasing amounts of information are being
generated and processed in the data centers of Amazon, Facebook,
Google, Dropbox, and many others. The demand for ever-increasing
amounts of cloud storage is supplied through the use of Distributed
Storage Systems (DSS), where data is stored on a network of
\textit{nodes} (hard drives and solid-state drives).

In the DSS paradigm, it is essential to store data redundantly, in
order to tolerate inevitable node
failures~\cite{BhaTatCheSavVoe04,GheGobLeu05,RheEatGeeWeaZhaKub03}.
Currently, the resilience against node failures is typically
the result of \emph{replication}, where several copies of each data
object are stored on different storage nodes. However, replication is
highly inefficient in terms of storage
capacity. Recently,~\emph{erasure-correcting codes} have been used in
DSS to reduce the large storage overhead of replicated
systems~\cite{DatOgg11,DimRamWuSuh11,HuaSimOguCalGopLiYek12}.

Apart from storage space, other metrics should be considered when
designing an actual DSS. However, in contrast with storage space,
these metrics are adversely affected by the straightforward use of
simple erasure-correcting codes.  One such metric is the {\em repair
bandwidth}: the amount of data that needs to be transferred when a
node has failed, and is thus replaced. This metric is highly relevant
as a prohibitively large fraction of the network bandwidth in a DSS
may be consumed by such repair operations. Let us term \emph{all} the
information stored by a DSS as \emph{the file}.  Traditional
erasure-correcting codes, and in particular~\emph{maximum distance
separable (MDS)} codes, usually require that \emph{all} the file be
downloaded in order to regenerate a failed node. Recently, Dimakis et
al.~\cite{DimGodWuWaiRam10} established a trade-off between the repair
bandwidth and the storage capacity of a node, and introduced a new
family of erasure-correcting codes, called \emph{regenerating codes},
which attain this trade-off. In particular, they proved that if a
\emph{large number} of storage nodes can be contacted during the
repair of a failed node, and only \emph{a fraction of their stored
data} is downloaded, then the repair bandwidth can be minimized.

Local repair of a DSS is an additional property which is highly
sought. The corresponding performance metric is termed the {\em locality}
of the coding scheme: the number of nodes that must
participate in a repair process when a particular node fails. Local
repair is of significant interest when a cost is associated
with contacting each node in the system. This is indeed the case in
real world scenarios, for example as the result of network
constraints.  Codes which enable local repairs of failed system nodes
are called \emph{locally repairable codes (LRCs)}.  These codes were
introduced by Gopalan et al.\ in~\cite{GopHuaSimYek12}. LRCs
which also minimize the repair bandwidth, called
codes with local regeneration, were considered
in~\cite{KamPraLalKum14,KamSilPraRawLalKoyKumVis13,RawKoySilVis14}.

Regenerating codes and LRCs are attractive
primarily for the storage of \emph{cold} data -- archival data that is
rarely accessed. On the other hand, they do not address the challenges
posed by the storage of frequently accessed \emph{hot} data.  For
example, hot-data storage must enable efficient reads of the same data
segments by several users \emph{in parallel}. This property is
referred to as \emph{availability}. Codes which provide both locality
and availability were first proposed in~\cite{RawPapDimVis16}.

Recently, codes with locality and availability have found another
application in the well known area of private information
retrieval~\cite{ChoGolKusSud98}. Shah, Rashmi, and
Ramchandran~\cite{ShaRasRam14} were the first to consider storage
overhead for this important concept.  In an important development,
Fazeli, Vardy, and Yaakobi~\cite{FazVarYaa15a,FazVarYaa15b}
demonstrated how codes with good availability can be used to save
storage and to obtain low storage overhead. Their new ideas have
motivated a series of papers with related results, e.g.,
\cite{BlaEtz16,BlaEtz17,FraGurWoo17,LinRos17,RaoVar16,VajRamKum17a,VajRamKum17b,ZhaWanWeiGe16}.
Other codes which were studied in the context of private information
retrieval are batch codes~\cite{IshKusOstSah04,AsiYaa17}. These codes also
have applications as distributed storage system codes~\cite{RawSonDimGaa16}.

Regenerating codes are described in terms of stored information in
nodes (servers). In other words, regenerating codes are usually array
codes~\cite{TamWanBru13}. Reconstructing the files and repairing
failed nodes are the main tasks of regenerating codes. LRCs
and codes with availability are usually described as
block codes, and access and/or repair is described in terms of symbols.

In this work we combine the two approaches and discuss two types of
locality (respectively, availability): node locality (availability),
which resembles the first approach, and symbol locality
(availability), which resembles the second approach. To our knowledge,
such a combined approach was not considered in the literature before.

Our solution approach will be based on array codes, constructed via
subspaces of a finite vector space. A subspace approach for DSS codes was
considered for the first time in~\cite{Hol13} and later
in~\cite{RavEtz15}. Our approach is slightly different from the
approach in these two papers. We shall employ spreads, $q$-Steiner
systems, and subspace transversal designs in our constructions. We
will also analyze the node and symbol, locality and availability, of
the resulting codes. This subspace approach for locality and availability is also novel.

\subsection{Our Contribution}
In this paper we present several constructions of array codes.  The
parameters of these codes are summarized in
Table~\ref{tb:tab_Summary}.  Note, that $\locs$ and $\locn$ denote
\emph{symbol locality} and \emph{node locality}, respectively, and
$\avs$ and $\avn$ denote the symbol availability and node availability,
respectively (for formal definitions see Definitions
\ref{def:array}-\ref{def:av} in the following section).

\begin{itemize}
\item
Construction \ref{con:allsub} is based on all the $b$-dimensional
subspaces of $\Fq^M$.  When $b=1$, it yields the classic simplex code,
and hence it can be considered as its generalization and $q$-analog.
\item
Construction \ref{con:onespread} is based on a $b$-spread of $\Fq^M$,
which are very important and well studied in projective geometry (see
the definition of a $b$-spread in Section~\ref{subsec:codes from
  subspace designs}).  This construction also yields the simplex code
when $b=1$, and when $M=2b$, it yields an MDS array code.  Moreover,
its dual code is a perfect array code (see
Lemma~\ref{lem:perfectbyte}).
\item
Construction \ref{con:allsub} and Construction \ref{con:onespread} are
based on the two extreme cases of the $q$-analog of combinatorial
designs. More generally, we provide Construction \ref{con:qsteiner},
which generalizes the previous two constructions.  It uses the
$q$-analog of block designs, namely, $q$-Steiner systems. However,
there is only one set of parameters (apart from the parameters of
Constructions \ref{con:allsub} and \ref{con:onespread}) where they are
known to exist. Nonetheless, it is conjectured that infinite families
of such designs exist (see Section~\ref{subsec:codes from subspace
  designs}).
\item
Construction \ref{con:std} is based on a subspace transversal
design. These designs have similar properties to the the ones of
$q$-Steiner systems, but unlike them, subspace transversal designs are
known to exist for many parameters (see the definition of a subspace
transversal design in Section~\ref{subsec:codes from subspace
  designs}).  In particular, we consider two types of constructions
from subspace transversal designs, namely
\begin{enumerate}
\item
based on a single parallel class of a subspace transversal design;
\item
based on all the subspaces in a subspace transversal design.
\end{enumerate}
When $M=2b$, the first construction produces an MDS array code.
In addition, the dual code of the code obtained from this construction
is an asymptotically perfect array code.
\end{itemize}

\begin{table}[htb]
\caption{Parameters of the constructed codes.}
\label{tb:tab_Summary}
\begin{center}
\begin{tabular}{l|l|l|l}
\textbf{Reference} & $\mathbf{[b\times n,M,d]}$ & \textbf{Symbol locality} & \textbf{Node locality} \\
\hline
Construction \ref{con:allsub} & $[b\times \sbinom{M}{b},M,q^{M-b}\sbinom{M-1}{b -1}]$ &
$\locs=\begin{cases}
  1 & 1<b<M,\\
  2 & b=1.
  \end{cases}$ &
  $\locn=2$
\\
Construction \ref{con:onespread} &
   $[b\times\frac{q^M-1}{q^b-1},M,q^{M-b}]$ &
    $\locs=2$&
    $2\leq \locn\leq b+1$ \\
Construction \ref{con:std}.1 &
$[b\times q^{M-b},M,q^{M-b}-q^{M-2b}]$&
$\locs=2$&
$\locn=\begin{cases}
  3 & q=2, \\
  2 & q>2.
  \end{cases}$\\
Construction \ref{con:std}.2&
 $[b\times q^{(M-b)t}, M, q^{(M-b)(t-1)}(q^{M-b}-q^{M-2b})]$&
$\locs=1$ &
$\locn\geq 2$\\
\end{tabular}
\end{center}
\end{table}

In addition to the node and symbol locality of the constructed codes
summarized in Table \ref{tb:tab_Summary}, we have node and symbol
availability for some of the codes. The code from Construction
\ref{con:allsub} has symbol availability
\[\avs=\begin{cases}
\sbinom{M-1}{b-1}-1 &  1<b<M,\\
\frac{q^{M-1}-1}{2} & b=1.
\end{cases}\]
and node availability
\[\avn=\begin{cases}
      \frac{1}{2}\parenv{\sbinom{M}{2}-1} &  2=b<M, \text{even } q\\
      \geq \frac{1}{2}\parenv{\sbinom{M}{2}-1-q(q^2+q-1)\sbinom{M-2}{2}} & 2=b<M,\text{odd } q.
      \end{cases}.\]
The symbol availability of the code from Construction \ref{con:std} (the one based on
all the subspaces in a subspace transversal design) is $\avs=q^{(M-b)(t-1)}-1$.

\subsection{Related Constructions}

Codes with locality $r$ and availability $t$ allow us to recover any
code symbol by using $t$ disjoint sets of cardinality $r$ (usually for
$r$ relatively small). This line of research has been extremely active
in the last few years as a consequence of its practical
importance. The results of some known code constructions with locality
and availability and their generalizations, mainly related to the
constructions presented in this paper, are summarized below.  We note
that our combined approach, that distinguishes between node and symbol
locality and availability, was not considered before. Many known
constructions in the literature are not array codes, therefore
precluding the distinction between nodes and symbols.  Thus, actual
comparison with previous works is mostly impossible, except for one
simple case mentioned below.

\begin{itemize}
\item
\textbf{Codes with locality and availability.}  Constructions of codes
with locality and availability were proposed
in~\cite{HuaYaaUchSie15,RawPapDimVis16,PamHolOgg13,TamBar14,WanZhaLiu15}.
Specifically, the construction presented
in~\cite{PamHolOgg13} is based on partial
geometries. Resolvable combinatorial designs, and modified pyramid
codes were used in \cite{RawPapDimVis16}. The approach
in~\cite{TamBar14} is based on orthogonal partitions and on
product codes. One-step majority-logic decodable codes and product
codes are used in~\cite{HuaYaaUchSie15}.
\item
\textbf{Codes with locality and availability over small fields.}
Codes over small alphabets (and in particular, binary codes) are of
particular interest due to their simple implementation.  The locality
properties of the family of binary simplex codes were proved
in~\cite{CadMaz15}.  Modifications of simplex codes based on anticodes
technique yield optimal codes with good locality and availability
properties, as shown in~\cite{SilZeh15}.  Binary cyclic LRCs were
considered in~\cite{GopCal14,ZehYaa15}.  Binary
codes for any given locality $r$ and availability $t$ are provided
in~\cite{WanZhaLiu15}.

\item
\textbf{Codes with local regeneration.} Codes that combine the
properties of LRCs with regenerating codes, by allowing to minimize
the repair bandwidth locally, were presented
in~\cite{KamPraLalKum14,KamSilPraRawLalKoyKumVis13,RawKoySilVis14}.
Most of these codes (i.e.,
\cite{KamSilPraRawLalKoyKumVis13,RawKoySilVis14}) are based on the
properties of linearlized polynomials.  To the best of our knowledge,
these are the only previously known \textit{array} codes that have
locality properties.  However, the locality for these codes is defined
only for nodes, and the symbol locality appears to be hard to extract
from the construction.

\item
\textbf{Other extensions and generalizations of LRCs.} Codes that
enable \textit{cooperative} local recovery from multiple erasures were
presented in~\cite{RawMazVis15}. In other words, these codes allow to
recover any small set of codeword symbols from a small number of other
symbols. Codes where symbols have \textit{different} localities were
considered in~\cite{ZehYaa16,KadSpr16}. Codes with
\textit{hierarchical locality}, which enable local recovery from
multiple erasures were presented in~\cite{SasAgaKum15}. The PIR array
codes considered in~\cite{BlaEtz16,BlaEtz17} have optimal symbol
availability, with symbol locality 2, for large number of nodes, but
their node locality and availability were not considered and again,
appear to be hard to extract.

\item
  \textbf{Fractional repetition codes.} Construction of such codes,
  e.g., in \cite{ElRRam10,KooGil11,ZhuShuLiHou14,SilEtz15}, provide
  arrays of repeating symbols. These were not intended originally for
  node and symbol locality and availability. However, their relatively
  simple structure allows us to find their parameters or bound them.
  In the notation of \cite{SilEtz15}, an $(n,\alpha,\rho)$-FR code
  (Fractional Repetition code) is composed of $\alpha\times n$ arrays
  with $\theta\eqdef n\alpha/\rho$ information symbols, each appearing
  in $\rho$ distinct columns. Thus, trivially, the symbol locality is
  $\locs=1$, the symbol availability is $\avs=\rho-1$. For nodes we
  have the trivial upper bounds of $\locn\leq \alpha$ and $\avn\leq
  \rho-1$. In \cite{SilEtz15} we find three constructions of FR codes:
  $[\alpha\times n, \alpha n/2,2]$ codes, $[\alpha\times \rho\alpha,
    \alpha^2, \rho]$ codes for $\rho\geq 3$, and
  $[(t+1)\times(s+1)(st+1),(t+1)(st+1),s+1]$ codes for $t\geq s$ (with
  further restrictions described in detail in
  \cite{SilEtz15}). However, the main disadvantage of these codes,
  compared with the codes we construct (see Table
  \ref{tb:tab_Summary}) is their low minimum distance.
\end{itemize}

\subsection{Paper Organization}

The rest of this paper is organized as follows. Preliminaries are
given in Section~\ref{sec:pre}. Our subspace approach, constructions
of codes, and analysis of their locality and availability, are
presented in Section~\ref{sec:subspaces}. We conclude in
Section~\ref{sec:conc} with a short discussion and some open problems.

%%%%%%%%%%%%%%%%%%%%%%%%%%%%%%%%%%%%%%%%%%%%%%%%%%%%%%%%%%%%%%%%%%%%%%%%
%%%%%%%%%%%%%%%%%%%%%%%%%%%%%%%%%%%%%%%%%%%%%%%%%%%%%%%%%%%%%%%%%%%%%%%%
%%%%%%%%%%%%%%%%%%%%%%%%%%%%%%%%%%%%%%%%%%%%%%%%%%%%%%%%%%%%%%%%%%%%%%%%
\section{Preliminaries}
\label{sec:pre}
Let $\F_q$ denote the finite field of size $q$. For a natural number
$m\in\N$, we use the notation $[m]\eqdef\mathset{1,2,\dots,m}$. We use
lower-case letters to denote scalars. Overlined letters denote
vectors, which by default are assumed to be column vectors.  Matrices
are denoted by upper-case letters.
% Most literature denotes codewords (which are usually vectors)
%by overlined lower-case letters.
However, the codewords of array codes, which are arrays (matrices), will
be denoted by bold lower-case letters. Thus, typically, we shall have
a generator matrix $G$, whose $j$th column is $\og_j$, and whose
$(i,j)$th entry is $g_{i,j}$. An array code will usually be denoted by
$C$, whose typical codeword will be denoted by $\bc$. We use $0$ to
denote the scalar zero, $\ozer$ for the all-zero column vector, and $\bzer$
for the all-zero matrix.  Also, given a (possibly empty) set of
vectors, $v_1,\dots,v_m\in\F_q^n$, their span is denoted by
$\spn{v_1,\dots,v_m}$.

Our main object of study is a linear array code, formally defined as
follows.
\begin{definition}
  \label{def:array}
 A $[b\times n,M,d]$ \emph{array code} over $\F_q$, denoted
$C$, is a linear subspace of $b\times n$ matrices over
$\F_q$. Matrices $\bc\in C$ are referred to as \emph{codewords}. The
elements of a codeword are denoted by $c_{i,j}$, $i\in[b]$, $j\in[n]$,
and are referred to as \emph{symbols}. Columns of codewords are
denoted by $\oc_j$, $j\in[n]$. We denote by $M\eqdef\dim(C)$ the
\emph{dimension} of the code as a linear space over $\F_q$. The
\emph{weight} of an array is defined as the number of non-zero
columns, i.e., for $\bc\in C$,
\[ \wt(\bc)\eqdef\abs{\mathset{ \oc_j ~:~ \oc_j\neq \ozer, j\in[n]}}.\]
Finally, the \emph{minimum distance} of the code, denoted $d$, is the
defined as the minimal weight of a non-zero codeword,
\[ d\eqdef\min_{\substack{\bc\in C \\ \bc\neq \bzer}} \wt(\bc).\]
\end{definition}

We make two observations to avoid confusion with other notions of
error-correcting codes. The first observation is that by reading the
symbols of codewords, column by column, and within each column, from
first to last entry, we may \textit{flatten} the $b\times n$ codewords
to vectors of length $bn$. This results in a code over $\F_q$ of
length $bn$, dimension $M$, but more often than not, a different
minimum distance, since the above definition considers non-zero
\emph{columns} and not non-zero symbols. Assume $G$ is an $M\times bn$
generator matrix for the \textit{flattened} code. By abuse of
notation, we shall also call $G$ the \emph{generator matrix} for the
original array code $C$. Note that in $G$, columns $(j-1)b+1,\dots,
jb$, correspond to the symbols appearing in the $j$th codeword column
in $C$. We shall call these $b$ columns in $G$ by the $j$th
\emph{thick column} of $G$, similarly to \cite{KamPraLalKum14}. Thus,
$G$ is a matrix comprised of $n$ thick columns, corresponding to the
$n$ columns of codewords in $C$.

\begin{example}
\label{ex:arrayCode}
Over $\F_2$, let $C$ be a $[2\times 5,5,3]$ array code, and let
\[\bc =\begin{pmatrix} 0&0&0&0&1 \\
      0&1&1&0&0
\end{pmatrix}\]
be a codeword of $C$ with weight 3. The corresponding flattened codeword
is $(0001010010)$, which is exactly the last row of the following
generator matrix $G$ for $C$,
\[G=\parenv{
      \begin{array}{c|c|c|c|c}
        10 & 00 & 01 & 01 & 00 \\
        00 & 10 & 00 & 01 & 01 \\
        01 & 00 & 10 & 00 & 01 \\
        01 & 01 & 00 & 10 & 00 \\
        00 & 01 & 01 & 00 & 10
      \end{array}
    },\]
which has 5 thick columns (separated by vertical lines).
\end{example}

The second observation is that we may use the well known isomorphism
$\F_q^b\cong \F_{q^b}$, and consider each column of a codeword as a
single element from $\F_{q^b}$. We get an $\F_q$-linear code over
$\F_{q^b}$ (sometimes called a \emph{vector-linear code}), of length
$n$, minimum distance $d$, but with a dimension (taken as usual over
$\F_{q^b}$) not necessarily $M$.

\begin{figure}[t]
\centering
\includegraphics[width=0.45\columnwidth]{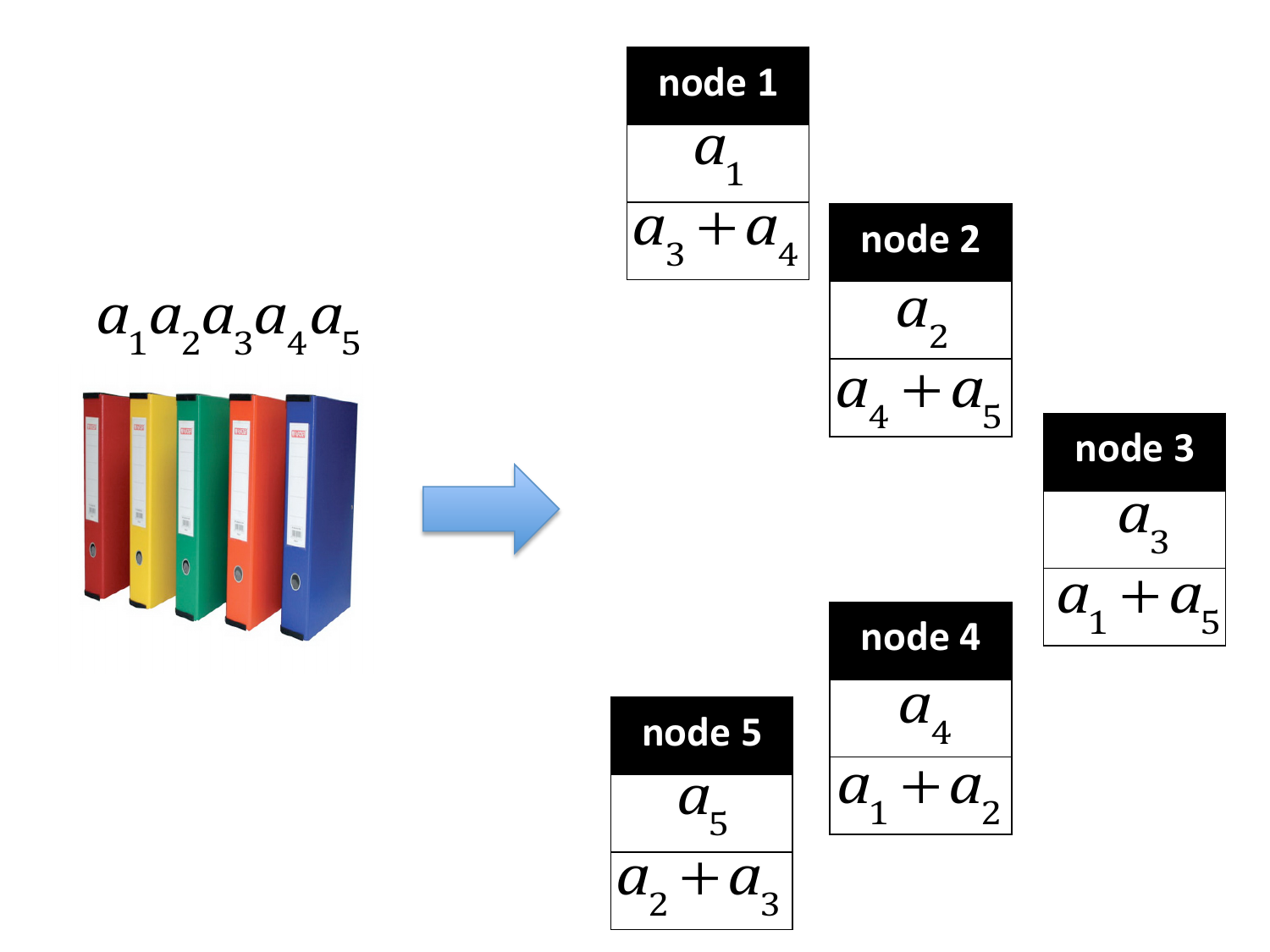}
\caption{Distributed storage system based on the binary $[2\times 5,5,3]$ array code from Example \ref{ex:arrayCode}.}
\label{fig:arrayCode}
\end{figure}

In a typical distributed-storage setup, we would like to store a file
containing $M$ sectors. We choose $\F_q$ such that it is large enough
to contain all possible sectors as symbols. The file is encoded into
an array $\bc\in C$ from a $[b\times n,M,d]$ array code.  Each
codeword column of $\bc$ is stored in a different node. The minimum
distance $d$ of the code ensures that any failure of at most $d-1$
nodes may be corrected. Figure \ref{fig:arrayCode} illustrates this
idea using the code from Example~\ref{ex:arrayCode}.

Two important properties of codes for distributed storage are
\emph{locality} and \emph{availability}. An important feature of this
paper is the distinction between \emph{symbol} locality and
\emph{node} locality (respectively, availability).  Note that this
approach is different from the standard one, where only node locality
and availability are considered.  The motivation to explore codes with
different types of locality and availability is the problem of
\textit{latent sector errors (LSEs)}, where individual sectors
(symbols) on a drive (node) become unavailable~\cite{SchDamGil10}. As
can be observed in the sequel, symbol locality can be smaller when
compared to the node locality. Thus, a more efficient recovery of a
single symbol is possible, compared with the recovery of an entire
node, since fewer nodes need to be contacted. Similarly, symbol
availability can be larger when compared to the node availability,
which also enhances the recovery process of a single symbol compared
with an entire node.

\begin{definition}
Let $C$ be a $[b\times n,M,d]$ array code. We say a codeword column
$j\in[n]$ has \emph{node locality $\locn$}, if its content may be
obtained via linear combinations of the contents of the recovery-set
columns. More precisely, there exists a recovery set
$S=\mathset{j_1,\dots,j_{\locn}}\subseteq[n]\setminus\mathset{j}$ of
$\locn$ other codeword columns, and scalars $a_{\ell,m}^{(i)}\in\F_q$,
$i,\ell\in[b]$, $m\in[\locn]$, such that for all $i\in[b]$,
\begin{equation}
  \label{eq:deflocn}
  c_{i,j} = \sum_{m=1}^{\locn} \sum_{\ell=1}^b a_{\ell,m}^{(i)} c_{\ell,j_{m}}
\end{equation}
simultaneously for all codewords $\bc\in C$. If all codeword columns
have this property, we say the code has node locality of $\locn$.

Similarly, we say the code has \emph{symbol locality $\locs$}, if for
every coordinate, $i\in[b]$ and $j\in[n]$, there exists a recovery set
$S=\mathset{j_1,\dots,j_{\locs}}\subseteq[n]\setminus\mathset{j}$ of
$\locs$ other codeword columns, and scalars $a_{\ell,m}\in\F_q$,
$\ell\in[b]$, $m\in[\locs]$, such that for every codeword $\bc\in C$,
\begin{equation}
  \label{eq:deflocs}
  c_{i,j} = \sum_{m=1}^{\locs} \sum_{\ell=1}^{b} a_{\ell,m} c_{\ell,j_m}.
\end{equation}
Thus, each code symbol may be recovered from the code symbols in
$\locs$ other codeword columns.
\end{definition}

Note that the coefficients in (\ref{eq:deflocs}) are not necessarily
the same as those in (\ref{eq:deflocn}). Additionally, it is obvious
that $\locs\leq \locn$.

Once locality is defined, we can also define availability.
\begin{definition}
  \label{def:av}
 The \emph{node availability}, denoted $\avn$, (respectively, the
 \emph{symbol availability}, denoted $\avs$) is the number of
 pairwise-disjoint recovery sets (as in the definition of locality)
 that exist for any codeword column (respectively, symbol). Note that
 each recovery set should be of size at most $\locn$
 (respectively,~$\locs$).
\end{definition}

\begin{example}
  One can verify that the code from Example \ref{ex:arrayCode} has
  symbol locality $\locs=2$, but node locality
  $\locn=3$. Additionally, it has symbol availability $\avs=2$, but
  node availability $\avn=1$.
\end{example}

We also recall some useful facts regarding Gaussian coefficients. Let
$V$ be a vector space of dimension $n$ over~$\F_q$. For any integer
$0\leq k\leq n$, we denote by $\sbinom{V}{k}$ the set of all
$k$-dimensional subspaces (\textit{$k$-subspaces}, in short) of $V$.
% also called \emph{a
%  Grassmannian}. The $q$-number of $k$ is defined as
%\[[k]_q\eqdef 1+q+q^2+\dots+q^{k-1}=\frac{q^k-1}{q-1},\]
%and by abuse of notation we denote
%\[[k]_q!\eqdef [k]_q[k-1]_q\dots [1]_q.\]
%
The \emph{Gaussian coefficient} is defined for $n$, $k$, and $q$ as
%\begin{align*}
%\sbinom{n}{k}_q&\eqdef \frac{[n]_q!}{[k]_q![n-k]_q!} \\
%&=\frac{(q^n-1)(q^{n-1}-1)\dots(q^{n-k+1}-1)}{(q^k-1)(q^{k-1}-1)\dots(q-1)}.
%\end{align*}
\begin{align*}
\sbinom{n}{k}_q&\eqdef\frac{(q^n-1)(q^{n-1}-1)\dots(q^{n-k+1}-1)}{(q^k-1)(q^{k-1}-1)\dots(q-1)}.
\end{align*}
Whenever the size of the field, $q$, is clear from the context, we
shall remove the subscript $q$.

It is well known that the number of $k$-subspaces of an
$n$-dimensional space over $\F_q$ is given by~$\sbinom{n}{k}$.  In a
more general form, the number of $k'$-subspaces of $V$
which intersect a given $k$-subspace of $V$ in an
$i$-subspace is given by
\begin{equation}
\label{eq:vecint}
q^{(k'-i)(k-i)}\sbinom{n-k}{k'-i}\sbinom{k}{i}.
\end{equation}
Additionally, the Gaussian coefficients satisfy the following
recursions,
\begin{align}
  \sbinom{n}{k} &= \sbinom{n-1}{k}+q^{n-k}\sbinom{n-1}{k-1} \nonumber \\
  &=  q^k\sbinom{n-1}{k}+\sbinom{n-1}{k-1}.
  \label{eq:gaussrec}
\end{align}
For more on Gaussian coefficients, the reader is referred to
\cite[Chapter 24]{LinWil01}.

%%%%%%%%%%%%%%%%%%%%%%%%%%%%%%%%%%%%%%%%%%%%%%%%%%%%%%%%%%%%%%%%%%%%%%%%
%%%%%%%%%%%%%%%%%%%%%%%%%%%%%%%%%%%%%%%%%%%%%%%%%%%%%%%%%%%%%%%%%%%%%%%%
%%%%%%%%%%%%%%%%%%%%%%%%%%%%%%%%%%%%%%%%%%%%%%%%%%%%%%%%%%%%%%%%%%%%%%%%

\section{A Subspace Approach to LRCs}
\label{sec:subspaces}

Let $C$ be a $[b\times n, M,d]$ array code over $\F_q$. Throughout
this section we further assume that $b\leq M$. We now describe an
approach to viewing such array codes which will lead to the main
results of this section.

Denote $V\eqdef\F_q^M$ the $M$-dimensional vector space over~$\F_q$.
Let $G$ be a generator matrix for the (flattened) array code
$C$. For each $j\in[n]$, we define $V_j$, such that  $V_j\in\bigcup_{k=0}^b
\sbinom{V}{k}$, to be the column space of the $j$th thick column of
$G$, i.e.,
\[ V_j \eqdef \spn{ \og_{(j-1)b+1,},\og_{(j-1)b+2},\dots,\og_{jb}}.\]
We say $V_j$ is associated with the $j$th thick column of~$G$, or
equivalently, associated with the $j$th column of the codewords of
$C$.

\begin{example}
The $2$-dimensional vector space associated with the second thick
column of the code from Example \ref{ex:arrayCode} is $V_2=
\spn{(01000)^T,(00011)^T}$.
\end{example}

The following equivalence is fundamental to the constructions
and analysis of this section.
\begin{lemma}
  Let $C$ be a $[b\times n,M,d]$ array code over $\F_q$, and let
  $V_j$, $j\in[n]$, be the subspaces associated with the codeword
  columns. Then
  $S=\mathset{j_1,\dots,j_m}\subseteq[n]\setminus\mathset{j}$ is a
  recovery set for codeword column $j\in [n]$, if and only if
  \[ V_j \subseteq V_{j_1}+V_{j_2}+\dots+V_{j_m}.\]
  Similarly, $S$ is a recovery set for symbol $(i,j)$, $i\in[b]$, if
  \[ \og_{(j-1)b+i} \in V_{j_1}+V_{j_2}+\dots+V_{j_m},\]
  where $\og_{(j-1)b+i}$ is the $i$th column in the $j$th thick column
  of a generating matrix $G$ for $C$.
\end{lemma}
\begin{IEEEproof}
  This is a simple restatement of \eqref{eq:deflocn} and \eqref{eq:deflocs}.
\end{IEEEproof}

With this equivalence, we may obtain the node/symbol
locality/availability using subspace properties of the thick columns
of a generating matrix.
Another definition of interest is the following.

\begin{definition}
  Let $C$ be a $[b\times n, M,d]$ array code over $\F_q$, and let
  $V_j$ be the subspace associated with the $j$th thick column.  If
  $\dim(V_j)=b$ for all $j\in[n]$ we call $C$ \emph{full column rank}.
\end{definition}

%%%%%%%%%%%%%%%%%%%%%%%%%%%%%%%%%%%%%%%%%%%%%%%%%%%%%%%%%%%%%%%%%%%%%%%%
%%%%%%%%%%%%%%%%%%%%%%%%%%%%%%%%%%%%%%%%%%%%%%%%%%%%%%%%%%%%%%%%%%%%%%%%
%%%%%%%%%%%%%%%%%%%%%%%%%%%%%%%%%%%%%%%%%%%%%%%%%%%%%%%%%%%%%%%%%%%%%%%%

\subsection{Generalized Simplex Codes via Subspaces}

We start with a construction of array codes which may be considered as
a generalization and a $q$-analog of the classical simplex code, the dual of the Hamming
code (see \cite[p.~30]{MacSlo78}).
\begin{construction}
  \label{con:allsub}
  Fix a finite field $\F_q$, positive integers ${1\leq b\leq M}$, and
  $V=\F_q^{M}$. Construct a $b\times \sbinom{M}{b}$ array code
  whose set of columns are associated with the subspaces
  $\sbinom{V}{b}$, each appearing exactly once. To make the dependence on the code parameters
    explicit, we denote this code by
    $C_b^M$.
\end{construction}

%We observe that the codes of Construction~\ref{con:allsub} form
%a generalization of simplex codes.
Note that when we choose $b=1$ in Construction \ref{con:allsub} we obtain
the simplex code. This fact will be used in the proof of Theorem~\ref{th:allsub} below.

We make a note here, which is also relevant for the constructions to
follow. Once we fix the set of subspaces associated with the codeword
columns, the code is constructed in the following way: for each
$j\in[n]$, and associated subspace $V_j$, we arbitrarily choose a set
of $b$ vectors from $\F_q^{M}$ that form a basis for $V_j$. These
$b$ vectors are placed (in some arbitrary order) as the columns
comprising the $j$th thick column of a generator matrix $G$. The
resulting matrix $G$ generates the constructed code\footnote{Permuting
the thick columns in the construction results in equivalent
codes. If a canonical representation is required, we may choose the
basis of each thick column to be in reduced row echelon form.}.

\begin{lemma}
\label{lem:auxnonker}
  Fix a finite field $\F_q$, positive integers $b<M$, and
  $V=\F_q^{M-1}$. For any $V'\in\sbinom{V}{b-1}$, given as the
  column space of an $(M-1)\times (b-1)$ matrix $G'$, and for any
  non-zero vector $\ou\in \F_q^{M-1}$ such that $\ou^T G'=\ozer^T$,
  the following hold:
  \begin{enumerate}
  \item
    If $\ox,\oy\in\F_q^{M-1}$ are in the same coset of $V'$, then
    $\ou^T \ox=\ou^T \oy$.
  \item
    The number of cosets of $V'$, all of whose vectors $\ox$ satisfy
    $\ou^T \ox= a$, for some fixed $a\in\F_q$, is exactly $q^{M-b-1}$.
  \end{enumerate}
\end{lemma}
\begin{IEEEproof}
  Denote the columns of $G'$ as $\og'_1,\dots,\og'_{b-1}$. If $\ox$
  and $\oy$ are in the same coset of $V'$, then there exist scalars
  $a_1,\dots,a_{b-1}$ such that
  \[ \ox = \oy + \sum_{j=1}^{b-1} a_j \og'_j.\]
  Multiplying on the left by $\ou^T$, and recalling that $\ou^T G'=\ozer^T$,
  we obtain the first claim.

  The number of cosets of $V'$ is exactly $q^{M-b}$, each containing
  $q^{b-1}$ vectors. Since $\ou\neq \ozer$, the number of vectors
  $\ox\in\F_q^{M-1}$ such that $\ou^T \ox=a$ is
  $q^{M-2}$. Dividing this by the number of vectors per coset we
  obtain the second claim.
\end{IEEEproof}

We are now ready for the first claim on the properties of the codes
from Construction \ref{con:allsub}.
\begin{theorem}
  \label{th:allsub}
  The array code obtained from Construction \ref{con:allsub} is a
  $[b\times \sbinom{M}{b},M,d]$ array code, with
  \[d=\sbinom{M}{b}-\sbinom{M-1}{b}=q^{M-b}\sbinom{M-1}{b -1}.\]
  Additionally, except for the all-zero array codeword, all other
  codewords have the same constant weight $d$.
\end{theorem}

\begin{IEEEproof}
  Apart from the minimum distance of the code, all other parameters
  are trivial. We shall prove the minimum distance property by proving
  the constant-weight property of the non-zero codewords by induction
  on $M$ and $b$ (we refer to this induction as \emph{induction A}).
  Additionally, we assert an auxiliary claim on the thick columns of
  the generator matrix, namely, that each thick column has rank
  $b$. We will prove this claim by induction as well (we refer to this
  second induction as \emph{induction B}).

  For the basis of induction A we have the following cases. When
  considering $C_M^M$, the codewords are $M\times 1$ arrays, and
  trivially, any non-zero codeword has weight
  \[ 1 = q^{M-M}\sbinom{M-1}{M-1}.\]
  Another base case is $C_1^M$. In the resulting generator matrix, each
  thick column contains just a single column, and the matrix is nothing
  but a generator matrix for the well known simplex code. The codewords
  are $1\times (q^M-1)/(q-1)$ arrays. The weight of the non-zero codewords
  in the simplex code is known to be $q^{M-1}$, and indeed we get a constant
  weight of
  \[ q^{M-1} = q^{M-1}\sbinom{M-1}{0}.\]
  We additionally note that in both cases, each thick column has rank
  $b$, i.e., the  basis for induction B holds.

  Assume now the claim holds for $C_{b-1}^{M-1}$ and for
  $C_{b}^{M-1}$, for both inductions, A and B. For the induction step
  we prove the claim also holds for $C_{b}^{M}$. Let their respective
  generating matrices be $G_{b-1}^{M-1}$ and $G_{b}^{M-1}$. Since we
  are not in any of the induction-base cases, we additionally have
  $1<b<M$.

  We construct a new matrix, $G$ by concatenating modified thick
  columns from $G_{b-1}^{M-1}$ and $G_{b}^{M-1}$.  We first take
  each thick column of $G_b^{M-1}$, append a bottom row of all
  zeros, and place it as a thick column of $G$. We call these columns
  \emph{thick columns of type I}.

  All the remaining thick columns of $G$, which we call of \emph{type
    II}, are formed by the thick columns of $G_{b-1}^{M-1}$ as
  follows. Consider such a single thick column, which is an
  $(M-1)\times (b-1)$ matrix on its own. Denote its column space by
  $V'\subseteq \F_q^{M-1}$, which by the hypothesis of induction B,
  has rank $b-1$. Thus, there are $q^{M-b}$ cosets of $V'$ in
  $\F_q^{M-1}$. Let $\ov'_1,\dots,\ov'_{q^{M-b}}$ be arbitrary coset
  representatives of the distinct cosets of $V'$. We create $q^{M-b}$
  thick columns in $G$ from the given thick column of $G_{b-1}^{M-1}$
  by placing it, each time with $\ov'_i$ as a $b$th column, and with
  an appended bottom row of $0,\dots,0,1$. In such thick columns of
  type II, the left $b-1$ coordinates are called \emph{the recursive
    part}, whereas the last coordinate is called \emph{the coset
    part}. The two types of thick columns of $G$ (depending on their
  source) are depicted in Figure \ref{fig:thicktype}.

  \begin{figure}[ht]
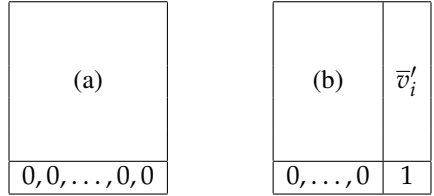

  \[ \begin{array}{|c|}
    \hline
    \\
    \\
    \text{(a)}\\
    \\
    \\
    \hline
    0,0,\dots,0,0\\
    \hline
  \end{array}
  \qquad\qquad
  \begin{array}{|c|c|}
    \hline
    &\\
    &\\
    \text{(b)} & \ov'_i\\
    &\\
    &\\
    \hline
    0,\dots,0 & 1\\
    \hline
  \end{array}
  \]
  \caption{The two types of thick columns in the constructed matrix
    $G$: a type I thick column, created by a thick column (a) from
    $G_{b}^{M-1}$, and a type II thick column, created by a thick
    column (b) from $G_{b-1}^{M-1}$ and one of its column-space
    coset representatives. }
  \label{fig:thicktype}
  \end{figure}

  Simple bookkeeping shows that we have $\sbinom{M-1}{b}$ thick
  columns of type I, and $q^{M-b}\sbinom{M-1}{b}$ thick columns of
  type II, for a total of
  \[ \sbinom{M-1}{b}+q^{M-b}\sbinom{M-1}{b-1} = \sbinom{M}{b}\]
  thick columns, where we used \eqref{eq:gaussrec}. They are easily
  seen to have distinct associated subspaces, each of dimension $b$,
  accounting for all the $b$-subspaces of $V=\F_q^{M}$. Thus, $G$ is
  indeed a generator matrix for the code from Construction
  \ref{con:allsub}, where each column has rank $b$.

  Now that we have proven a decomposition for the generator matrix
  $G$, we can proceed with the proof of the constant weight of all
  non-zero codewords. It is easily seen that $G$ has full rank. We
  consider several cases, depending on the rows of $G$ participating
  in the linear combination creating the codeword at question.

  In the simplest case, if a codeword of $C_b^M$ is formed by the
  last row of $G$ only, then its weight is
  $q^{M-b}\sbinom{M-1}{b-1}$, as the number of thick columns of
  type II.

  For the second case, let us consider a codeword $\bc\in C_b^M$
  formed by a linear combination of some rows from the first $M-1$
  rows of $G$. By the hypothesis of induction A, the thick columns of
  type I contribute $\sbinom{M-1}{b}-\sbinom{M-2}{b}$ to the weight of
  $\bc$. Also by the hypothesis of induction A, the recursive parts of
  thick columns of type II contribute
  $q^{M-b}(\sbinom{M-1}{b-1}-\sbinom{M-2}{b-1})$ to the
  weight. Finally, even if for some thick column of type II the
  recursive part may produce a combination of all zeros, the coset
  part may be non-zero, thus contributing to the weight of $\bc$.
  More precisely, we have $\sbinom{M-2}{b-1}$ recursive parts the
  linear combination zeros.  Therefore, by Lemma \ref{lem:auxnonker},
  the coset part of exactly $\sbinom{M-2}{b-1}(q-1)q^{M-b-1}$ becomes
  non-zero, and contributes to the weight of $\bc$. In total we get,
  \begin{align*}
    \wt(\bc) &= \sbinom {M-1}{b}-\sbinom{M-2}{b}\\
    &\quad\ +q^{M-b}\parenv{\sbinom{M-1}{b-1}-\sbinom{M-2}{b-1}}\\
    &\quad\ +\sbinom{M-2}{b-1}(q-1)q^{M-b-1}\\
    &=\sbinom{M}{b}-\sbinom{M-1}{b}.
  \end{align*}

  Finally, we consider a linear combination that, non-trivially, uses
  some rows from the set of $M-1$ first rows, as well as the last
  row. The $1$'s in the last row are located exactly at the coset part
  of thick columns of type II. Since by Lemma \ref{lem:auxnonker}, the
  linear combination results in an equal number of appearances of each
  element of $\F_q$ in the coset parts, an addition of a multiple of
  the last row will not change that, and the weight of the codeword
  remains the same as in the previous case.
\end{IEEEproof}

%We observe that the codes of Construction~\ref{con:allsub} form
%a generalization of simplex codes.
%When we choose $b=1$ in Construction \ref{con:allsub},
%the simplex code is obtained, a fact that was used in the proof of Theorem~\ref{th:allsub}.
\begin{lemma}
\label{lm:local A}
  The array code obtained from Construction \ref{con:allsub}, with
  parameters $b<M$, has node locality of $\locn=2$, and symbol
  locality of
  \[\locs=\begin{cases}
  1 & b>1,\\
  2 & b=1.
  \end{cases}\]
\end{lemma}
\begin{IEEEproof}
  Let $C$ be a code generated by Construction \ref{con:allsub} with a
  generator matrix $G$. We first examine the case of $b>1$. For symbol
  locality, given any column of $G$, denoted $\og\in\F_q^M$, by
  \eqref{eq:vecint}, there are exactly $\sbinom{M-1}{b-1}$
  $b$-subspaces of $\F_q^M$ containing $\og$, each
  corresponding to a thick column of $G$. Since $b<M$, we have
  $\sbinom{M-1}{b-1}>1$, and there exists a thick column different
  than the one containing the column $\og$, whose column space
  contains $\og$. Hence, $\locs=1$.

  For node locality, given any subspace $V_j$ associated with the
  $j$th thick column of $G$, we can easily find two other subspaces
  $V_{j_1}$ and $V_{j_2}$, $j\not\in\mathset{j_1,j_2}$, such that
  $V_{j}\subseteq V_{j_1}+V_{j_2}$. For example: fix a basis for
  $V_j$. Take the first basis element and complete it to a basis of
  some $b$-subspace of $\F_q^M$, denoted $V_{j_1}$. Take
  the remaining $b-1$ basis elements of $V_j$ and complete them to a
  different $b$-subspace, denoted $V_{j_2}$. This can
  always be done when $1<b<M$. Hence, $\locn=2$.

  Finally, we consider the case $b=1$. In this case, each thick column
  of $G$ comprises of a single column. By definition this means that
  $\locn=\locs$, and since each column may be shown as the sum of two
  other columns, we have $\locn=\locs=2$.
\end{IEEEproof}

We note that we ignored the case of $b=M$ in the previous lemma,
since then the array codewords have a single column, and locality is
not defined.

We now turn to consider availability. Symbol availability is trivial.
\begin{corollary}
\label{cor:avail A}
  The array code obtained from Construction~\ref{con:allsub}, with
  parameters $1<b<M$, has symbol availability
  \[\avs=\sbinom{M-1}{b-1}-1\]
  and for $b=1$ $\avs=\frac{q^{M-1}-1}{2}$.
\end{corollary}
\begin{IEEEproof}
  We use \eqref{eq:vecint} to find the number of associated subspaces
  containing a given vector.
\end{IEEEproof}

Unlike locality, it appears that determining the node availability is
a difficult task. We consider only the simplest non-trivial case of
$b=2$.
\begin{lemma}
\label{lm:avail A}
  \label{lem:availability_grassmannian}
  The array code obtained from Construction \ref{con:allsub}, with
  parameters $2=b<M$, has node availability
  \[ \avn=\frac{1}{2}\parenv{\sbinom{M}{2}-1},\]
  when $q$ is even, and
  \[ \avn\geq \frac{1}{2}\parenv{\sbinom{M}{2}-1-q(q^2+q-1)\sbinom{M-2}{2}},\]
  when $q$ is odd.
\end{lemma}

\begin{IEEEproof}
  Let us consider some codeword column of the code, and its associated
  subspace, $V=\spn{\ov_1,\ov_2}$. We count the number of
  pairwise-disjoint pairs of subspaces $U,W\neq V$, such that
  $V\subseteq U+W$.  We show how all subspaces (except for $V$) may be
  paired in such a manner, except perhaps for a few due to parity
  issues. We distinguish between two different kinds of subspaces,
  where the subspaces of the first kind intersect $V$ in a
  one-dimensional subspace (a projective point), and where the
  subspaces of the second kind have only trivial intersection with
  $V$.

  First, we consider subspaces of the first kind. There are
  $\sbinom{M-1}{1}-1=q\sbinom{M-2}{1}$ associated subspaces
  different form $V$ that contain a given vector $\ov\in V$, $\ov\neq
  \ozer$, and we denote them by $\cV_{\ov}$. Since there are
  $\sbinom{2}{1}=q+1$ projective points in $V$, denoted
  $\ov_1,\dots,\ov_{q+1}$, we have $q(q+1)\sbinom{M-2}{1}$
  associated subspaces which intersect $V$ in a one-dimensional
  subspace. Note that if $U\in\cV_{\ov_i}$ and $W\in\cV_{\ov_j}$, with
  $i\neq j$, then $V\subseteq U+W$.  We now further partition each
  $\cV_{\ov_i}$ into $q$ sets of equal size, arbitrarily. We denote
  these $\cV_{\ov_i}^j$, where $j\in[q+1]\setminus\mathset{i}$. The
  size of each such set is
  \[\abs{\cV_{\ov_i}^j} = \sbinom{M-2}{1}.\]
  Finally, for each $i,j\in[q+1]$, $i\neq j$, we arbitrarily create
  pairs of elements, one from $\cV_{\ov_i}^j$, and one from
  $\cV_{\ov_j}^i$. The total number of such pairs is
  $\binom{q+1}{2}\sbinom{M-2}{1}$.

  Next we consider associated subspaces of the second kind. There are
  $\sbinom{M}{2}-1-q(q+1)\sbinom{M-2}{1}$ such subspaces. We will
  prove that for even $q$ one can partition all these subspaces into
  disjoint pairs, and for odd $q$ one can partition all but a few such
  subspaces into disjoint pairs. The statement of the lemma then
  follows from this proof.

  Given an associated subspace $U=\spn{\ou_1,\ou_2}$, $U\cap
  V=\mathset{\ozer}$, we define a set $\cS_U$ of $q^{4}$ subspaces, as
  follows:
  \[\cS_U=\mathset{\spn{\ou_1+\ox_1,\ou_2+\ox_2} ~:~ \ox_1,\ox_2\in V}.
  \]
  Note that since $U\cap V=\mathset{\ozer}$, the vectors $\ou_1+\ox_1$ and
  $\ou_2+\ox_2$ are linearly independent. One can easily verify that
  $\cS_U$ is well defined, and the choice of two basis vectors,
  $\ou_1$ and $\ou_2$, does not change $\cS_U$.

  Additionally, if we have two distinct associated subspaces of the
  second kind, $U\neq U'$, then either $\cS_U\cap \cS_{U'}=\emptyset$
  or $\cS_U=\cS_{U'}$. To see that, assume $W_1\in \cS_U\cap\cS_{U'}$, i.e.,
  \begin{align*}
    W_1 &=\spn{\ou_1+\ox_1,\ou_2+\ox_2}\in \cS_U, \\
    W_1 &=\spn{\ou'_1+\ox'_1,\ou'_2+\ox'_2}\in \cS_{U'},
  \end{align*}
  with $\ox_1,\ox_2,\ox'_1,\ox'_2\in V$. Then there exist
  $\alpha_{1,1},\alpha_{1,2},\alpha_{2,1},\alpha_{2,2}\in\F_q$ such
  that
  \begin{align*}
    \ou_1+\ox_1 &= \alpha_{1,1}(\ou'_1+\ox'_1)+ \alpha_{1,2}(\ou'_2+\ox'_2),\\
    \ou_2+\ox_2 &= \alpha_{2,1}(\ou'_1+\ox'_1)+ \alpha_{2,2}(\ou'_2+\ox'_2),
  \end{align*}
  and
  \[ \Delta=\det\begin{pmatrix} \alpha_{1,1} & \alpha_{1,2} \\ \alpha_{2,1} & \alpha_{2,2}\end{pmatrix}\neq 0.\]
  We cannot have $\alpha_{1,1}=\alpha_{1,2}=0$, and we assume
  $\alpha_{1,2}\neq 0$ where the other case is symmetric. Then, given
  $W_2\in\cS_U$, $W_2=\spn{\ou_1+\oy_1,\ou_2+\oy_2}$, where
  $\oy_1,\oy_2\in V$, we define
  \begin{align*}
    \oy'_1 &\eqdef \ox'_1+\frac{\alpha_{1,2}}{\Delta}\parenv{\frac{\alpha_{2,2}}{\alpha_{1,2}}(\oy_1-\ox_1)-(\oy_2-\ox_2)},\\
    \oy'_2 &\eqdef \ox'_2+\frac{1}{\alpha_{1,2}}\parenv{\oy_1-\ox_1-\alpha_{1,1}(\oy'_1-\ox'_1)}.
  \end{align*}
  Obviously, $\oy'_1,\oy'_2\in V$. We also observe that
  \begin{align*}
    \ou_1+\oy_1 &= \alpha_{1,1}(\ou'_1+\oy'_1)+ \alpha_{1,2}(\ou'_2+\oy'_2),\\
    \ou_2+\oy_2 &= \alpha_{2,1}(\ou'_1+\oy'_1)+ \alpha_{2,2}(\ou'_2+\oy'_2),
  \end{align*}
  and so $W_2=\spn{\ou'_1+\oy'_1,\ou'_2+\oy'_2}\in\cS_{U'}$. Hence, if
  $\cS_{U}\cap\cS_{U'}\neq\emptyset$, then $\cS_U=\cS_{U'}$.

  Thus, as $U$ ranges over all associated subspaces of the second
  kind, $\cS_U$ partitions that set of subspaces into equivalence
  classes. We arbitrarily identify each such class with a subspace
  $U$, and a pair of basis vectors, $\ou_1,\ou_2\in U$.

  Depending on the parity of $q$ we have two cases. First we consider
  even $q$. We partition each class $\cS_U$, identified by $U$ and
  $\ou_1,\ou_2\in U$, into disjoint pairs as follows: We pair each
  \[W=\spn{\ou_1+\ox_1,\ou_2+\ox_2}\in\cS_U,\]
  with
  \[f(W)=\spn{\ou_1+\ox_1+\ov_1,\ou_2+\ox_2+\ov_2}\in\cS_U.\]
  Since $q$ is even, this is indeed well defined since
  $f(f(W))=W$. Additionally, the objective is met since
  \[ V=\spn{\ov_1,\ov_2}\subseteq W+f(W).\]

  When $q$ is odd, we partition each class $\cS_U$, identified by $U$ and
  $\ou_1,\ou_2\in U$, into disjoint pairs by pairing
  \[W=\spn{\ou_1+\ox_1,\ou_2+\ox_2}\in\cS_U,\]
  with
  \[f(W)=\spn{\ou_1-\ox_1,\ou_2-\ox_2}\in\cS_U.\]
  Except for $\ox_1=\ox_2=\ozer$, this is indeed a pairing since
  $f(f(W))=W$. Additionally, whenever $\ox_1$ and $\ox_2$ are linearly independent, we have
  \[ V=\spn{\ov_1,\ov_2}\subseteq W+f(W).\]
  The number of such pairs is $\frac{1}{2}(q^2-1)(q^2-q)$. Hence, we
  are not using $q(q^2+q-1)$ subspaces of the $q^4$ subspaces in
  $\cS_U$, and there are $\sbinom{M-2}{2}$ sets $\cS_U$.
\end{IEEEproof}

%%%%%%%%%%%%%%%%%%%%%%%%%%%%%%%%%%%%%%%%%%%%%%%%%%%%%%%%%%%%%%%%%%%%%%%%
%%%%%%%%%%%%%%%%%%%%%%%%%%%%%%%%%%%%%%%%%%%%%%%%%%%%%%%%%%%%%%%%%%%%%%%%
%%%%%%%%%%%%%%%%%%%%%%%%%%%%%%%%%%%%%%%%%%%%%%%%%%%%%%%%%%%%%%%%%%%%%%%%

\subsection{Codes from Subspace Designs}
\label{subsec:codes from subspace designs}

In this subsection we focus on constructing codes by using certain
subspace designs. We first present a different generalization of
simplex codes by using spreads. The resulting code is known, and we
analyze it for completeness, and for motivating another construction
that uses subspace designs.

Consider a finite field $\F_q$ and the vector space
$V\eqdef \F_q^M$. A \emph{$b$-spread of $V$} is a set
$\mathset{V_1,V_2,\dots,V_n}\subseteq\sbinom{M}{b}$ such that $V_i\cap
V_j=\mathset{\ozer}$ for all $i,j\in[n]$, $i\neq j$, and additionally,
$\bigcup_{i\in[n]} V_i = V = \F_q^M$.
Thus, except for the zero vector, $\ozer$, a spread is a partition of
$\F_q^M$ into subspaces. It is known that a $b$-spread exists if and
only if $b|M$. Simple counting shows that the number of subspaces in a
spread is
\[ n=\frac{q^M-1}{q^b-1}=\frac{\sbinom{M}{1}}{\sbinom{b}{1}}.\]
Let us start with a code obtained from a single spread. This code was
already described in~\cite{OggAnw11}, in the context of self-repairing codes, and we bring it here for
completeness.
\begin{construction}
  \label{con:onespread}
  Fix a finite field $\F_q$, positive integers $b|M$, and
  $V=\F_q^M$. Construct a $b\times \sbinom{M}{1}/\sbinom{b}{1}$ array
  code whose set of columns are associated with the subspaces of a
  $b$-spread of $V$, each appearing exactly once.
\end{construction}

\begin{theorem}
  \label{th:onespread}
  The array code obtained from Construction \ref{con:onespread} is a
  $[b\times\sbinom{M}{1}/\sbinom{b}{1},M,q^{M-b}]$ array
  code. Additionally, except for the all-zero array codeword, all
  other codewords have the same constant weight.
\end{theorem}
\begin{IEEEproof}
  Denote $u\eqdef\sbinom{M}{1}/\sbinom{b}{1}$. Consider an $M\times
  bu$ generator matrix $G$ for the code $C$ from Construction
  \ref{con:onespread}. It contains $u$ thick columns, each made up of
  $b$ columns. Let $G_i$, $i\in[u]$, be the $M\times b$ submatrix of
  $G$ containing the $b$ columns of the $i$th thick column, i.e., $G=(
  G_1 | G_2 | \dots | G_u)$.

  We now take each $G_i$, $i\in[u]$, and construct from it an $M\times
  (q^b-1)$ matrix we call $G^\ext_i$, whose columns are the column
  space of $G_i$ except for $\ozer$. We concatenate those to obtain
  the $M\times (q^M-1)$ matrix
  \[ G^\ext\eqdef \parenv{ G^\ext_1 | G^\ext_2 | \dots | G^\ext_u}.\]
  Since the thick columns of $G$ form a $b$-spread of $\F_q^M$, the
  columns of $G^\ext$ contain each possible vector exactly once,
  except for $\ozer$.

  We now observe that a row of $G^\ext_i$ is $\ozer^T$ iff it is
  $\ozer^T$ in $G_i$. Additionally, a non-zero row of $G^\ext_i$
  contains exactly $q^{b-1}$ occurrences of each non-zero element of
  $\F_q$. Finally, each non-zero element of $\F_q$ appears $q^{M-1}$
  times in each row of $G^\ext$. Thus, given a row of $G^\ext$,
  exactly $q^{M-1}/q^{b-1}=q^{M-b}$ of its $u$ thick columns are
  non-zero, implying the same for the corresponding row in $G$, and
  then the associated array codeword has weight $q^{M-b}$.

We now want to prove the same thing for every non-trivial linear
combination of the rows of $G$. First, note that having a $b$-spread
of $\F_q^M$ is equivalent to having $\rank(G_i)=b$, and $\rank(G_i |
G_j)=2b$, for all $i,j\in[u]$, $i\neq j$. Consider a linear
combination of rows $i_1,i_2,\dots,i_{\ell}$ of $G$, each with a
non-zero coefficient, resulting in a row vector $\ov^T$. Replace row
$i_\ell$ of $G$ by the vector $\ov^T$ to obtain a new matrix $G'=(G'_1
| G'_2 | \dots | G'_u)$. Since the rank is invariant to such
operations, $\rank(G'_i)=b$ and $\rank(G'_i|G'_j)=2b$ for all
$i,j\in[u]$, $i\neq j$. Thus, $G'$ is equivalent to a $b$-spread
(perhaps different from the original one induced by $G$). Using
the same logic as before, exactly $q^{M-b}$ of the thick columns
of $\ov^T$ are non-zero, completing the proof.
\end{IEEEproof}

\begin{lemma}
The array code obtained from Construction \ref{con:onespread}, $b<M$,
has symbol locality $\locs=2$, and its node locality satisfies $2\leq
\locn\leq b+1$. Moreover, there exist such array codes with $\locn\leq
M/b$.
\end{lemma}
\begin{IEEEproof}
To prove the symbol locality, we note that any column of $G$ can be
presented as a linear combination of two other columns which belong to
two other distinct thick columns. Otherwise, if these two columns
belong to the same thick column, we obtain a contradiction to the
definition of a spread. Thus, $\locs\leq 2$. We also obviously have
$\locs\geq 2$, otherwise we contradict the partitioning property of
the spread.

For the node locality, since in general $\locs\leq \locn$ we have that
$2\leq \locn$. Let $\mathset{\ov_1,\dots,\ov_b}$ be a basis for a
thick column of $G$ which represents an element (subspace) $V_i$ of
the spread. Take an arbitrary $\ow\not\in V_i$ and define $\ou_i\eqdef
\ov_i+\ow$, for all $i\in[b]$. Observe that $\ow$ and all the vectors
$\ou_i$, $i\in[b]$, belong to $b+1$ different subspaces (corresponding
to thick columns) in a spread, or else these would intersect $V_i$
non-trivially. Clearly, $V_i$ can be reconstructed from these $b+1$
subspaces.

For the remainder of the proof let us assume that the spread is
constructed in a specific way, inferred from \cite{EtzSil09}, given in
more detail in \cite{GabPil11}, and described as follows. Every
element (subspace) in the constructed spread is presented as the row
space of a row-reduced echelon-form $b\times M$ matrix $(\bzer | \bzer
  | \dots | \bzer | I_b | A_1 | A_2 | \dots | A_t)$, where each block
is of size $b\times b$, $I_b$ is the $b\times b$, identity matrix, and
$(A_1| \dots| A_t)$ is a codeword of a Gabidulin code of length $bt$
and minimum rank distance $b$. Of particular interest are the ``unit''
subspaces,
\[ U_i \eqdef \rowsp (\underbrace{\bzer | \dots | \bzer}_{i-1} | I_b | \bzer | \dots | \bzer),\]
for all $i\in[M/b]$.  Obviously,
\[ \sum_{i=1}^{M/b} U_i = \F_q^M.\]
Thus, except for unit subspaces from $U\eqdef
\mathset{U_i}_{i\in[M/b]}$, for every other subspace of the spread,
the set $U$ is a recovery set of $M/b$ thick columns.

We are left with the task of finding recovery sets of unit subspaces
of the form $U_i$. For every $i\in[M/b-1]$, we have
\[ U_i \subseteq U_{i+1} + \rowsp(\underbrace{\bzer|\dots | \bzer}_{i-1} | I_b | A | \bzer | \dots | \bzer),\]
where $A\neq \bzer$ is a codeword of the above-mentioned Gabidulin
code. Finally,
\[ U_{M/b} \subseteq U_{M/b-1}+\rowsp(\bzer|\dots | \bzer | I_b | A ),\]
since $A$ is full rank due to the minimum rank distance of the
Gabidulin code. Thus, each $U_i$ has a recovery set of size $2\leq
M/b$.
\end{IEEEproof}

The code of Construction \ref{con:onespread} is also a generalization
of the simplex code. Indeed, when we take $b=1$ the resulting
generator matrix is that of a simplex code.
\begin{corollary}
  When $M=2b$, the code from Construction \ref{con:onespread} is
  an MDS array code with $\locn=\locs=2$.
\end{corollary}
\begin{IEEEproof}
  The node and symbol locality are trivial since the subspaces
  associated with thick columns have a pair-wise trivial intersection,
  and therefore the sum of any two such subspaces gives the entire
  space since $M=2b$. The code is MDS since it is a $[b\times (q^b+1),
    2b, q^b]$ array code.
\end{IEEEproof}

Up to this point we constructed codes by specifying their generator
matrix.  We now turn to consider their dual codes by reversing the
roles of generator and parity-check matrices.
We first require the following simple lemma.

\begin{lemma}
  \label{lem:dual}
  Let $C$ be a $[b\times n, M, d]$ array code over $\F_q$ that is full
  column rank. If the size of the smallest recovery set for a symbol
  of $C$ is of size $\ell$, then the dual code, $C^\perp$, is a
  $[b\times n, bn-M,\ell+1]$ array code. In particular, if the symbol
  locality of \emph{every symbol} of $C$ is $\locs$, then $C^\perp$ is
  a $[b\times n, bn-M, \locs+1]$ array code.
\end{lemma}
\begin{IEEEproof}
  Let $G$ be a generator matrix for $C$. The smallest recovery set of
  size $\ell$ together with the full column rank property imply that
  the smallest set of linearly dependent columns of $G$ includes
  columns from exactly $\ell+1$ thick columns. Considering $G$ as a
  parity-check matrix for $C^\perp$, we obtain that the any non-zero
  codeword of $C^\perp$ has at least $\ell+1$ non-zero columns. The
  rest of the code parameters are trivially obtained.
\end{IEEEproof}

The dual code of the code from Construction \ref{con:allsub} has a
small distance $d=2$, and is therefore not very interesting. However,
the code from Construction \ref{con:onespread} presents a more
interesting situation.
\begin{lemma}
  \label{lem:perfectbyte}
  Let $C$ be a code from Construction \ref{con:onespread}. Then its
  dual, $C^\perp$, is a $[b\times \sbinom{M}{1}/\sbinom{b}{1},
    b\sbinom{M}{1}/\sbinom{b}{1}-M, 3 ]$ array code. Additionally,
  $C^\perp$ is a perfect array code.
\end{lemma}
\begin{IEEEproof}
  The minimum distance follows from Lemma \ref{lem:dual} since the
  locality of all symbols in $C$ is $2$. To show that $C^\perp$ is
  perfect, note that the ball of radius $1$ has size
  \[ \Phi_1\eqdef 1+\frac{\sbinom{M}{1}}{\sbinom{b}{1}}(q^b-1)=q^{M}.\]
  Hence,
  \[\abs{C^\perp}\cdot\Phi_1 =q^{b\sbinom{M}{1}/\sbinom{b}{1}},\]
  which is equal to the size of the entire space.
\end{IEEEproof}

We note that the code of Lemma \ref{lem:perfectbyte} has already been
described as a perfect byte-correcting code in \cite{HonPat72,Etz98}.

At this point we stop to reflect back on Construction \ref{con:allsub}
and Construction \ref{con:onespread}. We contend that the two are in
fact two extremes of a more general construction using the $q$-analog
of Steiner systems.

\begin{definition}
  \label{def:qsteiner}
  Let $F_q$ be a finite field. A \emph{$q$-analog of a Steiner system}
  (a \emph{$q$-Steiner system} for short), denoted $S_q[t,k,n]$, is a
  set of subspaces, $\cB\subseteq\sbinom{\F_q^n}{k}$, such that every
  subspace from $\sbinom{\F_q^n}{t}$ is contained in exactly one
  element of $\cB$.
\end{definition}

In light of Definition~\ref{def:qsteiner}, we note that the subspaces
associated with the columns of Construction \ref{con:allsub} form a
$q$-Steiner system $S_q[b,b,M]$. Similarly, the subspaces associated
with the columns of Construction~\ref{con:onespread} form a
$q$-Steiner system $S_q[1,b,M]$. Both are therefore extreme (and
trivial) cases of a more general construction we now describe.

\begin{construction}
  \label{con:qsteiner}
  Fix a finite field $\F_q$, and let $\cB\subseteq\sbinom{\F_q^M}{b}$
  be a $q$-Steiner system $S_q[t,b,M]$. Construct an array code whose
  set of columns are associated with the subspace set $\cB$, each
  appearing exactly once.
\end{construction}

The main problem with the approach of Construction \ref{con:qsteiner}
is the fact that we need a $q$-Steiner system. Such systems are
extremely hard to find \cite{SchEtz02,BraEtzOstVarWas16}, with the
only known ones, different $S_2[2,3,13]$, found by computer
search~\cite{BraEtzOstVarWas16}. But, there is still a potential in this
construction as it is believed that infinite families of $q$-Steiner systems exist~\cite{BraEtzOstVarWas16}.

An alternative approach uses a structure that is ``almost'' a
$q$-Steiner system, and is more readily available -- a subspace
transversal design (see \cite{EtzSil13}).

\begin{definition}
Let $\F_q$ be a finite field. A \emph{subspace transversal design} of
group size $q^m=q^{n-k}$, block dimension $k$, and strength $t$, denoted
by $\std_q(t,k,m)$ is a triple $(\cV,\cG,\cB)$, where
\begin{enumerate}
\item
    $\cV\eqdef\sbinom{\F_q^n}{1}\setminus\cV_0^{(n,k)}$, called
    \emph{the points}, where $\cV_0^{(n,k)}$ is defined to be the set
    of all $1$-subspaces of $\F_q^n$ all of whose vectors
    start with $k$ zeros, and where $\abs{\cV}=\sbinom{k}{1}q^m$.
\item
    $\cG$ is a partition of $\cV$ into $\sbinom{k}{1}$ classes of size
    $q^m$, called \emph{the groups}.
\item
    $\cB\subseteq\sbinom{\F_q^n}{k}$, called \emph{the blocks}, is a
    collection of subspaces that contain only points from $\cV$, with
    ${\abs{\cB}=q^{mt}}$.
\item
    Each block meets each group in exactly one point.
\item
    Each $t$-subspace of $\F_q^n$, with points only from
    $\cV$, which meets each group in at most one point, is contained
    in exactly one block.
\end{enumerate}
  An $\std_q(t,k,m)=(\cV,\cG,\cB)$ is called \emph{resolvable} if the
  set~$\cB$ may be partitioned into sets $\cB_1,\dots,\cB_s$, called
  \emph{parallel classes}, where each point is contained in exactly
  one block of each parallel class $\cB_i$.
\end{definition}

Unlike $q$-Steiner systems, subspace transversal designs are known to exist in a
wide range of parameters, as shown in the following
theorem~\cite{EtzSil13}.

\begin{theorem} \cite[Th.~7]{EtzSil13}
  \label{th:resolv}
  For any $1\leq t\leq k\leq m$, and any finite field $\F_q$, there
  exists a resolvable $\std_q(t,k,m)=(\cV,\cG,\cB)$, where the block
  set $\cB$ may be partitioned into $q^{m(t-1)}$ parallel classes,
  each one of size $q^m$, such that each point is contained in exactly
  one block of each parallel class.
\end{theorem}

%We are now ready for the next construction.

\begin{construction}
  \label{con:std}
  Fix a finite field $\F_q$, $M\geq 2b$, and let $(\cV,\cG,\cB)$ be a
  $\std_q(t,b,M-b)$ with parallel classes
  $\cB_1,\cB_2,\dots,\cB_s$. Construct the following two array codes:
  \begin{itemize}
  \item
    An array code $\Cpar$ whose set of columns are associated with the
    subspaces in a single parallel class, $\cB_i$, each appearing
    exactly once.
  \item
    An array code $C$ whose set of columns are associated with the
    subspaces in $\cB$, each appearing exactly once.
  \end{itemize}
\end{construction}

The code $\Cpar$ is in fact an auxiliary code we shall use to prove
the parameters of the code $C$, and is perhaps of interest on its own.

\begin{theorem}
  \label{th:cpar}
  Let $\Cpar$ be the code from Construction \ref{con:std}. Then
  $\Cpar$ is a $[b\times q^{M-b},M,q^{M-b}-q^{M-2b}]$ array code, with
  $2^b-1$ codewords of full weight $q^{M-b}$, and the other non-zero
  codewords of weight $q^{M-b}-q^{M-2b}$.  Moreover, the symbol
  locality of $\Cpar$ is $\locs=2$, and its node locality is
  \[\locn=\begin{cases}
  3 & q=2, \\
  2 & q>2.
  \end{cases}\]
\end{theorem}

\begin{IEEEproof}
  The size and dimension of the array code follow from Theorem
  \ref{th:resolv}. The rest of the proof follows the same logic as the
  proof of Theorem \ref{th:onespread}.

  Denote $u\eqdef q^{M-b}$. Consider an $M\times bu$ generator matrix
  $G$ for $\Cpar$. It contains $u$ thick columns, each made up of $b$
  columns.  Let $G_i$, $i\in[u]$, be the $M\times b$ submatrix of $G$
  containing the $b$ columns of the $i$th thick column, i.e.,
  $G=(G_1|G_2|\dots|G_u)$.

  We now take each $G_i$, $i\in[u]$, and construct from it an $M\times
  (q^b-1)$ matrix we call $G^\ext_i$, whose columns are the column
  space of $G_i$ except for $\ozer$. We concatenate those to obtain
  the $M\times u(q^b-1)$ matrix
  \[G^\ext \eqdef (G^\ext_1|G^\ext_2|\dots|G^\ext_u).\]
  Since we used a single parallel class, the columns of $G^\ext$
  contain each possible vector exactly once, except for columns
  beginning with $b$ zeros. In other words, the subspaces of dimension
  $b$ that correspond to the thick columns of $G$, together with the
  subspace of dimension $M-b$ of all vectors starting with $b$ zeros,
  form a partition of the non-zero vectors of $\F_q^M$.

  We now observe that a row of $G^\ext_i$ is $\ozer^T$ iff it is
  $\ozer^T$ in $G_i$. Additionally, a non-zero row of $G^\ext_i$
  contains exactly $q^{b-1}$ occurrences of each non-zero element of
  $\F_q$. It is now a matter of simple counting, to obtain that each
  of the first $b$ rows of $G^\ext$ has all of its $u=q^{M-b}$ thick
  columns non-zero, and the remaining lower $M-b$ rows of $G^\ext$
  have exactly $q^{M-b}-q^{M-2b}$ non-zero thick columns in each row.

  Finally, consider a linear combination of the rows of $G$ that
  involves rows $i_1,i_2,\dots,i_\ell$, all with non-zero
  coefficients, and resulting in a row $\ov^T$. As in the proof of
  Theorem \ref{th:onespread}, let us replace row $i_\ell$ of $G$ with
  $\ov^T$ to obtain a new generator matrix $G'$. Again, the subspaces
  the correspond to the thick columns of $G'$ induce a partition of
  the non-zero vectors of $\F_q^M$ into subspaces of dimension $b$ and
  a single subspace of dimension $M-b$. Therefore, we conclude that
  the resulting row corresponds to an array codeword of weight either
  $q^{M-b}$ or $q^{M-b}-q^{M-2b}$ depending on whether
  $i_1,\dots,i_\ell\in[b]$ or not. This gives us a total of $q^b-1$
  codewords in $\Cpar$ of weight $q^{M-b}$, and the remaining non-zero
  codewords of weight $q^{M-b}-q^{M-2b}$.

  To complete the proof, the symbol locality is $\locs=2$, since any
  column of $G$ may be easily be given as a sum of two other columns
  of $G$ (which must also reside in distinct thick columns), due to
  the partition of $\F_q^M$ discussed above. To prove the node
  locality we recall that any thick column of $G$ corresponds to a
  lifted MRD codeword, i.e., $(I_b|A)^T$, where $A$ is a codeword of
  a linear MRD code of dimension $M-b$.  When $q=2$, we can recover
  $(I_b|A)^T$ by noting that
  \[ (I_b|A)^T = (I_b|A')^T + (I_b|A+A')^T + (I_b|\bzer)^T,\]
  where $A'$ is a codeword of the lifted MRD code, $A'\neq A$, and
  where we use the fact that $M-b\geq 2$. When $q>2$, let
  $\alpha\in\F_q$, $\alpha\neq 0,1$. Then we can recover $(I_b|A)^T)$
  by noting that
  \[ (I_b|A)^T = \alpha^{-1}(I_b|\alpha A) + (\alpha-1)\alpha^{-1}(I_b|\bzer)^T,\]
  thus proving $\locn=2$ for $q>2$.
\end{IEEEproof}

\begin{corollary}
  When $M=2b$, the code $\Cpar$ from Construction \ref{con:std} is an
  MDS array code with $\locn=\locs=2$.
\end{corollary}
\begin{IEEEproof}
  The node and symbol locality are trivial since the subspaces
  associated with thick columns have a pair-wise trivial intersection,
  and therefore the sum of any two such subspaces gives the entire
  space since $M=2b$. The code is MDS since it is a $[b\times q^b,
    2b, q^b-1]$ array code.
\end{IEEEproof}

\begin{corollary}
  Let $\Cpar$ be the code from Construction \ref{con:std}. Then its
  dual code, $\Cpar^\perp$ is a $[b\times q^{M-b},bq^{M-b}-M, 3]$
  array code that is asymptotically perfect.
\end{corollary}

\begin{IEEEproof}
  The parameters of the code follow from Lemma \ref{lem:dual} and from
  the proof of Theorem \ref{th:cpar}. Note that the size of a ball
  of radius $1$ is equal to
  \[ \Phi_1\eqdef 1+q^{M-b}(q^b-1).\]
  The size of the entire space is $q^{bq^{M-b}}$. Then
  \begin{align*}
    \frac{\abs{\Cpar^\perp}\cdot\abs{\Phi_1}}{q^{bq^{M-b}}}&=\frac{q^{bq^{M-b}-M}(1+q^{M-b}(q^b-1))}{q^{bq^{M-b}}}\\
    &=\frac{1+q^{M}-q^{M-b}}{q^{M}}=1+q^{-M}-q^{-b},
  \end{align*}
  and this ratio tends to $1$ when $b,M\to\infty$, implying the code
  family is asymptotically perfect.
\end{IEEEproof}

 \begin{example}
 \label{ex:liftedGab}
   Let $b=3$, $M=6$, $q=2$. A generator matrix $G$ for the $[3\times 8,
     6, 7]$ MDS array code $\Cpar$ from Construction \ref{con:std} is
   given by
   \[
   G=\parenv{
    \begin{array}{c|c|c|c|c|c|c|c}
      100 & 100 & 100 & 100 & 100 & 100 & 100 & 100 \\
      010 & 010 & 010 & 010 & 010 & 010 & 010 & 010 \\
      001 & 001 & 001 & 001 & 001 & 001 & 001 & 001 \\
      000 & 100 & 001 & 010 & 101 & 011 & 111 & 110 \\
      000 & 010 & 101 & 011 & 111 & 110 & 100 & 001 \\
      000 & 001 & 010 & 101 & 011 & 111 & 110 & 100
    \end{array}
  }.
  \]
 \end{example}

We now move on to examine the second code of Construction
\ref{con:std}. To avoid degenerate cases, we consider only $t\geq 2$.

\begin{theorem}
  Let $C$ be the code from Construction \ref{con:std}, with $t\geq
  2$. Then $C$ is a $[b\times q^{(M-b)t}, M, d]$ array code
  \[ d= q^{(M-b)(t-1)}(q^{M-b}-q^{M-2b}).\]
  The symbol and node locality of the code satisfy $\locs=1$, and
  $\locn\geq 2$. Its symbol availability is $\avs=q^{(M-b)(t-1)}-1$.
\end{theorem}

\begin{IEEEproof}
  The codeword size, as well as the minimum distance follow
  immediately by noting that there are $q^{(M-b)(t-1)}$ parallel
  classes, and a generator matrix for $C$ is simply the concatenation
  of generators for $\Cpar$ (for each of the parallel classes). The
  minimum distance then follows from Theorem \ref{th:cpar}.

  Additionally, each point (i.e., a column of $G$) is contained
  exactly once in each of the $q^{(M-b)(t-1)}$ parallel classes in a
  single subspace (i.e., the column span of a thick column of
  $G$). Thus, as long as $t\geq 2$, the symbol locality is $\locs=1$,
  and the availability is $\avs=q^{(M-b)(t-1)}-1$. Trivially, by the
  properties of the subspace transversal design, no subspace
  associated with a thick column appears twice, and hence $\locn\geq
  2$.
\end{IEEEproof}

\section{Conclusion}
\label{sec:conc}

We have suggested the usage of codes based on subspaces for the
purpose of locality and availability in distributed storage codes. We
introduced the concepts of symbol locality and symbol availability in
addition to the known node locality and node availability. We
constructed generalized simplex codes and Hamming codes from subspaces
and subspace designs (including $q$-Steiner systems, and subspace
transversal designs). We have found some of their locality and
availability parameters, or bounded them. In addition to the unsolved
questions in this paper, this topic has many more directions for
future research, e.g.:

\begin{enumerate}
\item Find new codes and designs, based on subspaces, with good locality
and availability properties.

\item Find upper bounds on the symbol locality and availability for
  codes based on subspaces and find codes which attain these bounds.

\item Develop the theory of PIR codes based on subspaces and find such
  good codes which outperform the known codes.
\end{enumerate}

\bibliographystyle{IEEEtranS}
\bibliography{allbib}
\end{document}